\documentstyle[aps,preprint,epsfig,eqsecnum]{revtex}
\begin{document}
\draft
\title{Consistent description of NN and $\pi$N interactions using the
solitary boson exchange potential}
\author{L. J\"ade}
\address{Theoretische Kernphysik, Universit\"at Hamburg\\
Luruper Chaussee 149, D-22761 Hamburg}
\date{\today}
\maketitle
\begin{abstract}
A unified description of NN and $\pi$N elastic 
scattering is presented in the
framework of the one solitary boson exchange potential (OSBEP). 
This model already successfully applied to analyze NN scattering is now
extended to describe $\pi$N scattering while also improving its accuracy
in the NN domain. We demonstrate the importance of 
regularization of $\pi$N
scattering amplitudes involving $\Delta$ isobars and derivative
meson-nucleon couplings, as this model 
always yields finite amplitudes 
without recourse to phenomenological form factors. 
We find an empirical scaling relation of the meson self 
interaction coupling constants consistent with that previously found in
the study of NN scattering. 
Finally, we 
demonstrate that the OSBEP model does not contradict the soft-pion
theorems of $\pi$N scattering. 
\end{abstract}
\narrowtext
\section{Introduction}
In low and medium energy hadron-hadron physics, the NN and
$\pi$N interactions play a most significant role in our  
understanding of the dynamics of strong
interactions. In that energy regime, specific effects of Quantum
Chromodynamics (QCD) are hidden in
effective degrees of freedom such as baryon and meson fields. 
There are a
number of models which address this problem, of which there exist QCD
inspired approaches which aim is to connect phenomena to fundamental
QCD features, such as chiral symmetry. 
With this approach, NN \cite{SKY62,WAM94,ORD94,SHA95} and
$\pi$N \cite{WEI67,BER97} interactions have been sought with means 
that led 
to fundamental insights like the $\pi$N low-energy theorems. But to
date, none of these QCD inspired models 
give an accuracy in description of experimental data
as that provided by phenomenological boson exchange
\cite{nipot,papot,Mach87,LEE91,PJ91,GRO93,SDHS94} or
inversion potential investigations \cite{KOH94,SAN97a,BEC95,SAN97b}.
It seems that chiral symmetry is not a 
dominant factor in the NN and $\pi$N scattering observables 
up to elastic
threshold \cite{hhi}. 
Nonetheless, even a chiral symmetry breaking
phenomenological approach should be based on concepts which eventually 
connect to a chiral symmetry maintaining model. As our analysis
will show, chiral symmetry is restored in the boson exchange model in
the limit $m_{\pi}\to 0$. 
    
A useful exercise is to interpolate between the QCD inspired and the
accurate phenomenological hadron-hadron interaction models. We do so  
using the one solitary boson exchange potential
OSBEP in application to NN scattering \cite{jae97}. The basic idea is to
parametrize the effects of chiral symmetry via nonlinear terms in the
meson Lagrangian with a structure  
equivalent to the linear $\sigma$-model. In 
contrast to that linear $\sigma$-model, we do not impose symmetry
conditions on coupling constants and masses. Rather we take these
entities as free parameters so 
breaking chiral symmetry to some extent. Then, we solve
the decoupled nonlinear meson field equations analytically and quantize
the quasiclassical solutions, so defining {\it solitary mesons}, 
to obtain
the propagator. Proper normalization of the meson
fields then ensures that all self-energy diagrams remain finite. 
A notable result of this approach is that the coupling constants and
masses obey an empirical scaling law; one which is similar to the 
symmetry
constraints of chiral models. A hope is that this scaling law can be
related to some underlying symmetry eventually.
   
The OSBEP model has been determined successfully in $np$ and $pp$ 
interactions \cite{jae98}. Thus we now seek its extension to provide  
a consistent model for both NN and $\pi$N interactions. 
To do so, we must include the $\Delta$ isobar specifically and use a 
chiral symmetry conserving PV
$\pi$NN meson-baryon coupling rather than the PS coupling used
previously \cite{jae98}. As a consequence, 
the proper normalization in the OSBEP model has to be adjusted to
provide finite results for
the self-energy diagrams involving $\Delta$ isobars and derivative
meson-baryon vertices. The refit of the parameters which enter the
NN potential even yields 
an improvement to the quality of the original model fit. But the major
achievement is that we now have a 
unified framework for both NN and $\pi$N interactions. 
   
Consistent application of a potential model in NN as well as $\pi$N
interactions has long been an unresolved puzzle. The major concern 
is that the $\pi$NN form factor differed in analyses of these
systems. The NN data demand a rather hard cut-off mass 
(e.\,g.\ $\Lambda_{\pi NN}=1.7\,$GeV in the Bonn-B potential 
\cite{Mach89}),
and that value can not be reconciled with the much softer cut-off
necessary to fit $\pi$N data below 1\,GeV. 
Sch\"utz {\it et al.} \cite{SDHS94} conjecture the
reason to be that the form factors 
depend on all momenta in the external legs of the vertices. 
However, Holzwarth and Machleidt \cite{HM97} state that it is
impossible to describe NN and $\pi$N interactions consistently if one
uses an analytical parameterization like the monopole form of most boson
exchange models, and instead propose a Skyrme-model form factor which
might be appropriate for both systems. However, to our knowledge, 
no potential model involving a form factor
parameterization exists to date that not only sensibly describes both
systems but also gives a sufficiently accurate fit to scattering data. 
  
As indicated above, we have found a 
unified NN and $\pi$N potential built upon the OSBEP model for the form
factors. Since all self-energy diagrams are regularized {\it by
construction} in this model, it is corollary that 
finite scattering amplitudes result for both scattering systems. More
important, fits to NN as well as $\pi$N data have given a consistent
parameter set which provides an agreement with the data as good as found
using the best conventional and separate 
NN or $\pi$N models. A notable feature of our result is the 
low number of parameters we have to specify. There are no adjustable
cut-off masses in our model and many possible parameters of the meson
nonlinearities are not as they are interrelated 
by a simple scaling relation which leaves the pion self-interaction
coupling constant only as adjustable parameter. 
Besides that, the only parameters we have, therefore, 
are the meson-meson and meson-baryon coupling
constants which are not fixed by experiment or symmetry relations.
  
After we give an overview 
of the main concepts of the OSBEP model in Sec.\ \ref{osbep}, we show in
Sec.\ \ref{propnorm} how the proper
normalization serves to yield finite scattering amplitudes, therein we
introduce also the scaling law for the masses and self-interaction 
coupling
constants. After that, we briefly sketch the application in NN 
scattering in
Sec.\ \ref{NN}. Finally, in Sec.\ \ref{pin}, 
we present the $\pi$N solitary boson exchange
potential and in Sec.\ \ref{res} we compare the results for NN and
$\pi$N phase shifts given by this unified model with those obtained
using conventional models and empirical phase shift analyses. 
Additionally, we address the problem of chiral symmetry by calculating
scattering lengths as function of the pion mass and show that our model
does not contradict the Weinberg-Tomozawa relations. 
\section{The OSBEP model}
\label{osbep}
Motivated by the linear $\sigma$-model approach \cite{Itzz80}, we
assume that nonlinear self-interactions for each meson field
($\beta=\pi,\eta,\rho,\omega,\sigma,\delta$) entering the boson exchange
potential lead to meson Lagrangians
\begin{equation}
{\cal L}_{\beta} = \frac{1}{2}\left(\partial_{\mu}\Phi_{\beta}
\partial^{\mu}\Phi_{\beta}-m_{\beta}^2\Phi_{\beta}^2\right)
-\frac{\lambda^{\beta}_1}{2p+2}\Phi_{\beta}^{2p+2}-
\frac{\lambda^{\beta}_2}{4p+2}
\Phi_{\beta}^{4p+2}+{\cal L}_{int}.
\label{ml}
\end{equation}
Note that by choosing proper values for $\lambda^{\beta}_1$,
$\lambda^{\beta}_2$ and $p$, one can retain the structure
of the linear $\sigma$-model. For convenience, spin and isospin indices
are dropped. The Lagrangian ${\cal L}_{int}$ contains meson-meson as
well as meson-baryon vertices which enter the NN and $\pi$N scattering
amplitudes and will be discussed below.
     
The main
assumption now is that as $t\to\pm\infty$, 
the meson fields only decouple
from {\it external} sources and thus the nonlinear
self-interaction current has to appear in the field equation for each
Fourier component  
\begin{equation}
\partial_{\mu}\partial^{\mu}\Phi_{\beta}(x,k)
+m_{\beta}^2\Phi_{\beta}(x,k)
+\lambda^{\beta}_1\Phi_{\beta}^{2p+1}(x,k)+
\lambda^{\beta}_2
\Phi^{4p+1}(x,k)=0.
\label{fe}
\end{equation}
Quasiclassical solutions can be obtained by the method of base functions
\cite{Burt81}. Essentially, one makes the ansatz $\Phi=\Phi(\varphi)$,
where $\varphi$ are free wave solutions of the Klein-Gordon
equation. This reduces Eq.\ (\ref{fe}) to an ordinary differential
equation which is solved by direct integration and the solutions
can be expressed as a power series in $\varphi$. 
As a naive quantization rule, inserting free wave operators
for $\varphi(x,k)$ gives the {\it solitary meson fields} with 
\begin{equation}
\label{sol}
\Phi_{\beta}(x,k)=
\sum_{n=0}^{\infty}C_n^{1/{2p}}(w_{\beta})\;b_{\beta}^n\;
\varphi_{\beta}^{2pn+1}(x,k),
\end{equation}
where
\begin{equation}
\label{fs}
\varphi_{\beta}(x,k)\equiv\frac{1}{\sqrt{2\omega_kVD_k^{(\beta)}}}
\;a_{\beta}(k)\;
e^{-ikx}.
\end{equation}
Here, $V$ is the volume of the system and
$\omega_k=(\vec{k}\,^2+m_{\beta}^2)^{1/2}$. 
Note that we have used a factor $1/\sqrt{D_k^{(\beta)}}$, 
a Lorentz-invariant function of $k$, which is of use in the 
proper normalization of the solitary meson fields that we give later in
Sec.\ \ref{propnorm}. The coefficients $C^a_n(x)$ are Gegenbauer
polynomials and $w_{\beta}$ and $b_{\beta}$ are
functions of the coupling constants and the order $p$ of the
self-interaction, namely 
\begin{equation}
w_{\beta} = \displaystyle{\frac{1}{b_{\beta}}
\frac{\lambda^{\beta}_1}{4(p+1)m_{\beta}^2}}\,,
\label{bw}
\end{equation}
where
\begin{equation}
b_{\beta} = 
\sqrt{\left(\displaystyle{\frac{\lambda^{\beta}_1}
{4(p+1)m_{\beta}^2}}\right)^2
-\displaystyle{\frac{\lambda^{\beta}_2}
{4(2p+1)m_{\beta}^2}}}\,.
\end{equation}
After solution of the field equations the interactions 
between mesons and baryons are treated perturbatively, and we assume
that ${\cal L}_{int}$ contains the couplings:
\begin{itemize}
\item Scalar meson-baryon coupling ($\beta=\sigma,\delta$)
\begin{equation}
\label{smbb}
{\cal L}_{mbb}^{(s)}(x)=-g_{\beta}
:\bar{\Psi}(x)\Psi(x):\tilde{\Phi}_{\beta}(x),
\end{equation}
\item Pseudovector meson-baryon coupling ($\beta=\pi,\eta$) 
\begin{equation}
\label{pvmbb}
{\cal L}_{mbb}^{(pv)}(x)=\frac{g_{\beta}}{2M}
:\bar{\Psi}(x)\gamma_5\gamma_{\mu}\Psi(x):\partial^{\mu}
\tilde{\Phi}_{\beta}(x),
\end{equation}
\item Vector meson-baryon coupling ($\beta=\rho,\omega$)
\begin{equation}
\label{vmbb}
{\cal L}_{mbb}^{(v)}(x)=g_{\beta}
:\bar{\Psi}(x)\left[\gamma_{\mu}\tilde{\Phi}_{\beta}^{\mu}(x)
+\frac{\kappa g_{\beta}}{2M}\sigma_{\mu\nu}
\partial^{\mu}\tilde{\Phi}_{\beta}^{\nu}(x)\right]\Psi(x):,
\end{equation}
where
$$
\sigma_{\mu\nu}=\frac{i}{2}[\gamma_{\mu},\gamma_{\nu}],
$$
\item $\pi N\Delta$ coupling
\begin{equation}
\label{pnd}
{\cal L}_{\pi N\Delta}(x)=\frac{g_{\pi
N\Delta}}{m_{\pi}}\bar{\Psi}(x)\vec{\hat{T}}\left(x_{\Delta}
\gamma^{\mu}\gamma_{\nu}\Psi^{\nu}_{\Delta}(x)+
\Psi^{\mu}_{\Delta}(x)\right)
\partial_{\mu}\vec{\tilde{\Phi}}_{\pi}(x)+\mbox{h.\,c.},
\end{equation}
\item $\sigma\pi\pi$ coupling
\begin{equation}
\label{spp}
{\cal L}_{\sigma\pi\pi}(x)=
\frac{g_{\sigma\pi\pi}}{2m_{\pi}}
\tilde{\Phi}_{\sigma}(x)\partial_{\mu}\vec{\tilde{\Phi}}_{\pi}(x)
\partial^{\mu}\vec{\tilde{\Phi}}_{\pi}(x),
\end{equation}
\item $\rho\pi\pi$ coupling
\begin{equation}
\label{rpp}
{\cal
L}_{\rho\pi\pi}(x)=g_{\rho\pi\pi}\vec{\tilde{\Phi}}^{\mu}_{\rho}(x)\Big(
\vec{\tilde{\Phi}}_{\pi}(x)
\times\partial_{\mu}\vec{\tilde{\Phi}}_{\pi}(x)
\Big).
\end{equation}
\end{itemize}
In these couplings $\Psi(x)$ are nucleon isospinors and, 
for isovector mesons, the operator $\tilde{\Phi}_{\beta}(x)$ has to be
replaced by $\vec{\tau}\vec{\tilde{\Phi}}_{\beta}(x)$.
To avoid double counting, 
the vertex operator is a weighted projection of the sum over all
Fourier components of the solitary meson field 
given in Eq.\ (\ref{sol}), i.\,e.\
\begin{equation}
\label{vo}
\tilde{\Phi}_{\beta}(x)=\sum_{N,N^{'},\vec{k}}
\frac{1}{\sqrt{N^{'}!}}|N^{'},k\rangle\langle
N^{'},k|\left(\Phi_{\beta}(x,k)+\Phi^{\dagger}_{\beta}(x,k)\right) 
|N,k\rangle\langle N,k|\frac{1}{\sqrt{N!}}\,.
\end{equation}
Since the propagator is used as the probability for a solitary meson to
move between the interaction vertices $x$ and $y$, we must 
define it using the fields defined in Eq.\ (\ref{vo}) by  
\begin{equation}
\label{propdef}
iP_{\beta}(x-y)=\langle 0|T\;\tilde{\Phi}(x)\tilde{\Phi}(y)|0\rangle.
\end{equation}
Inserting Eq.\ (\ref{sol}) into Eq.\ (\ref{vo}) one obtains the momentum
space amplitude \cite{jae97},
\begin{equation}
iP_{\beta}(k^2,m_{\beta})=
\sum_{n=0}^{\infty}\Big[C_n^{1/{2p}}(w_{\beta})\Big]^2
\frac{b_{\beta}^{2n}}{(2V)^{2pn}}\frac{(2pn+1)^{2pn-2}}
{{D_{k,n}^{(\beta)}}^{2pn+1}(\vec{k}\,^2+M_{n,\beta}^2)^{pn}}
\;i\Delta_F(k^2,M_{n,\beta}),
\label{pprop1}
\end{equation}
with the Feynman propagator being
\begin{equation}
\label{fprop}
i\Delta_F(k^2,M_{n,\beta})=\frac{i}{k^2-M_{n,\beta}^2}\ ,
\end{equation}
and a mass spectrum given by 
$$
M_{n,\beta}=(2pn+1)m_{\beta}\ .
$$
The normalization, $D_{k,n}^{(\beta)}$ in the
propagator Eq.\ (\ref{pprop1}), is obtained from the normalization
$D_k^{(\beta)}$ in Eq.\ (\ref{fs}) by substituting
$$
k^{\mu}\quad\to\quad\frac{1}{2pn+1}\;k^{\mu}.
$$
It is useful at this point to introduce 
dimensionless coupling constants $\alpha_{\beta}$,
$\alpha_1^{\beta}$ and $\alpha_2^{\beta}$ respectively
\begin{eqnarray}
\alpha_{\beta} & = & 
\displaystyle{\frac{b_{\beta}}{(2m_{\beta}V)^p}}\ , 
\nonumber\\
 & & \nonumber\\
\alpha_1^{\beta} & = & \frac{\lambda^{\beta}_1}
{4(p+1)m_{\beta}^2(2m_{\beta}V)^p}\ , \nonumber\\
 & & \nonumber\\
\alpha_2^{\beta} & = & \frac{\lambda^{\beta}_2}
{4(2p+1)m_{\beta}^2(2m_{\beta}V)^{2p}}\ .
\label{a1a2}
\end{eqnarray}
The final amplitude, which we define to be the {\it solitary
meson propagator}, then is
\begin{equation}
\label{pprop}
iP_{\beta}(k^2,m_{\beta})=\sum_{n=0}^{\infty}
\Big[C_n^{1/{2p}}(w_{\beta})\Big]^2
\frac{(m_{\beta}^p\alpha_{\beta})^{2n} 
(2pn+1)^{2pn-2}}{{D_{k,n}^{(\beta)}}^{2pn+1}(\vec{k}\,^2
+M_{n,\beta}^2)^{pn}}\;i\Delta_F(k^2,M_{n,\beta}),
\end{equation}
with
\begin{equation}
\label{wa}
w_{\beta}=\frac{\alpha_1^{\beta}}{\sqrt{{\alpha_1^{\beta}}^2-
\alpha_2^{\beta}}}\ .
\end{equation}
For $p=1/2$, one gets the amplitude for 
scalar fields while with $p=1$ the amplitude is that for pseudoscalar
particles. Vector
mesons require $p$ to be 1 and each term of the sum is
multiplied with a Minkowski tensor,
\begin{equation}
\label{mt}
f^{\mu\nu}_n=\left(-g^{\mu\nu}+
\frac{k^{\mu}k^{\nu}}{M_{n,v}^2}\right).
\end{equation}
\section{Proper normalization}
\label{propnorm}
\begin{figure}[t]\centering
\begin{picture}(5.0,11.0)(2.5,1.5)
\epsfig{figure=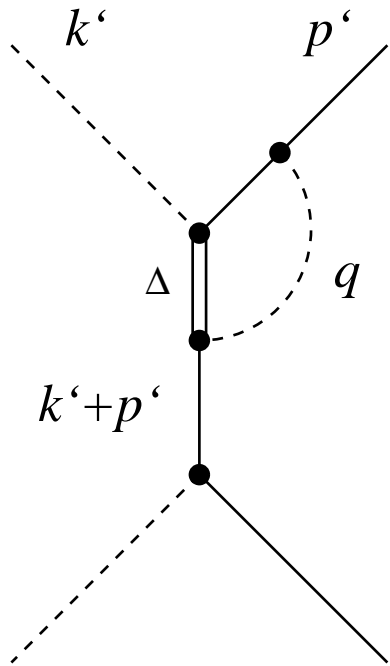,width=5.0cm}
\end{picture}
\caption[vertex correction]{$\pi N\Delta$ vertex correction.} 
\label{pinse}
\end{figure}

The proper normalization constant in Eq.\ (\ref{pprop}) is now fixed by
physical boundary conditions. First, we impose the constraints familiar
from renormalization theory, i.\,e.\ the propagator has to have a pole
of residue $i$ at the on-shell point, $k^2=m_{\beta}^2$. In addition,
we assume that (i) all amplitudes are Lorentz invariant, (ii)
$D_k^{(\beta)}$ is dimensionless and larger than unity, (iii) the
fields vanish when the interaction vanishes and, most important,
(iv) all self-energy diagrams are finite. This leads to the ansatz
\begin{eqnarray}
\label{dk}
D_{k}^{(\beta)} & = & \left\{1+\left[\left(
\frac{1}{\alpha_1^{\beta}4(p+1)(2m_{\beta})^p}\right)^{\frac{2}{p}}
+\left(\frac{1}{\alpha_2^{\beta}4(2p+1)
(2m_{\beta})^{2p}}\right)^{\frac{1}{p}}
\right]\right.\\
 & & \nonumber\\
 & & \left.\times
\left(\displaystyle{\sqrt{\vec{k}\,^2+m_{\beta}^2}-k_0}\right)^2
\right\}^{N_{pn}^{(\beta)}}.\nonumber
\end{eqnarray}
In the case that there is only one nonlinear term in the Lagrangian
($\alpha_2^{\beta}\equiv 0$), the term in Eq.\ \ref{dk} 
containing $\alpha_2^{\beta}$ is to be deleted. 
The exponent $N_{pn}^{(\beta)}$ can be chosen for each
meson type to yield finite scattering amplitudes. 
At this point, it is crucial to note that assuming a momentum-dependent
normalization $D_k^{(\beta)}$ for the fields in Eq.\ (\ref{sol}) {\it
always} gives finite self-scattering amplitudes for all 
interactions. Even in cases where the interaction is non-renormalizable
in standard models, such as with massive 
spin-1, $\rho$ and $\omega$ mesons,
our method can be applied. On the other hand, an energy-dependent
normalization affects the canonical commutation relations in coordinate
and momentum space. The equal-time 
commutator of the field operator and
its conjugate momentum is no longer a $\delta$-function
$\delta(\vec{x}-\vec{x}\,^{'})$ in
coordinate space, as in conventional field theoretical models, but 
approaches a {\it finite} value for $\vec{x}=\vec{x}\,^{'}$ and vanishes
otherwise. This can be interpreted as a finite particle size due to the
self-interaction. Another important point of the model is that
causality is preserved since 
the equal-time commutators of the fields remain unchanged, i.\,e.\
$$
[\Phi_{\beta}(\vec{x},t;k),\Phi_{\beta}(\vec{x}\,^{'},t;k)]
=[\tilde{\Phi}_{\beta}(\vec{x},t),
\tilde{\Phi}_{\beta}(\vec{x}\,^{'},t)]=0.
$$
To determine $N_{pn}^{(\beta)}$ one has to consider 
the most divergent self-energy amplitude for each type of meson. 
For scalar and vector
mesons, this is the first correction to the two-point function, Eq.\ 
(\ref{propdef}). Whereas for scalar mesons it is sufficient to choose
$$
D_k^{(s)}={\cal O}(k^2)\qquad\Rightarrow\qquad N_{pn}^{(s)}=1,
$$
for vector mesons one has to use
$$
D_k^{(v)}={\cal O}(k^4)\qquad\Rightarrow\qquad N_{pn}^{(v)}=2,
$$
due to the additional momentum dependence in Eq.\ (\ref{mt}). For the
pion however, the $\pi N\Delta$ vertex correction 
(see Fig.\ \ref{pinse}) 
is the most divergent amplitude. The combination of derivative coupling
in Eq.\ (\ref{pnd}) and the
$\Delta$-propagator which grows linearly with momentum, requires a
strong normalization for the pion. Therefore we have to choose
$$
D_k^{(ps)}={\cal O}(k^6)\qquad\Rightarrow\qquad N_{pn}^{(ps)}=3,
$$
to obtain finite results for all self-energy diagrams in NN {\it and}
$\pi$N interactions. The meson masses and normalizations used are listed
in Table \ref{mm}. Note that in our former work \cite{jae97,jae98}, 
$N_{pn}^{(ps)}=1$ sufficed since we used PS coupling for the $\pi$NN and
$\eta$NN vertex and the $\Delta$-isobar was not treated. This
modification now demands a refit of the parameters entering the NN
potential to maintain (or improve upon) the accuracy of fits to data. 
\section{Application to NN interactions}
\label{NN}
The concept of the NN solitary boson exchange potential is very similar
to the Bonn-B OBEP \cite{Mach89}. It has been described in detail in
Refs.\ \cite{jae97} and \cite{jae98}. The inclusion of the 
$\Delta$-isobar
and the PV coupling for the $\pi$NN and $\eta$NN vertices do not
significantly  
change the actual form of the potential. The $\Delta$ intermediate 
states do not contribute in the one boson exchange approximation and 
the PV coupling on-shell is identical to, and off-shell is very similar 
to, the PS coupling when the potential
is evaluated in the Blankenbecler-Sugar (BbS) reduction of the
Bethe-Salpeter equation \cite{Mach87}. 
Negative energy states do not contribute. 
We note also that Machleidt \cite{Mach89} has shown that, by slightly
changing the parameters, it is possible to
obtain equally good results for the Bonn-B potential using either the
PS or PV coupling. We confirm this
result and obtain fits of similar quality with both PS and PV coupling. 
The most important modification in the NN potential 
is to change the proper normalization exponent, $N_{pn}^{(\beta)}$,
of the pseudoscalar mesons ($\pi$ and $\eta$). 
The question arises whether the empirical
scaling relation for the self-interaction coupling constants and masses,
found in a previous comparison of the solitary meson propagator to the
Bonn-B form factors \cite{jae97}, remains valid. Using the strong
normalization, $N_{pn}^{(\beta)}=3$ for $\pi$ and $\eta$, a similar
analysis indicates that the scaling relation generalizes to
\begin{equation}
\label{scaling}
\frac{\alpha_{\beta}}{\sqrt{N_{pn}^{(\beta)}}}=
\frac{\alpha_{\pi}}{\sqrt{N_{pn}^{(\pi)}}}
\left(\frac{m_{\pi}}{m_{\beta}}\right)^p,
\end{equation}
and thus still serves to minimize the number of parameters we need to
specify. Note that we
simplified the model by setting $\alpha_2^{\beta}=0$ so that 
$\alpha_{\beta}=\alpha_1^{\beta}$ is the only self-interaction
parameter for all mesons. In the linear
$\sigma$ model, which motivated our ansatz for the nonlinear terms in
the meson Lagrangian, chiral symmetry also demands $\alpha_2^{\beta}=0$ 
for pseudoscalar and vector mesons. On the other hand, this does not
apply for scalar mesons. However, the scalar $\sigma$ meson in the
potential model serves as a parameterization of two-pion-exchange. It 
is not a fundamental particle as considered in the $\sigma$ model. The
second scalar meson, the $\delta$, only contributes little. 
\section{The $\pi$N solitary boson exchange potential}
\label{pin}
The structure of the $\pi$N boson exchange potential was adopted from
the work of Pearce and Jennings \cite{PJ91}. 
The only changes arise for the form factors which 
can be dropped due to the proper normalization of the solitary meson
fields and the three-dimensional reduction of the scattering equation  
to account for solitary mesons in the intermediate $\pi$N
states. 
  
\begin{figure}[t] \centering
\begin{picture}(8,9)(-0.7,0.5)
\epsfig{figure=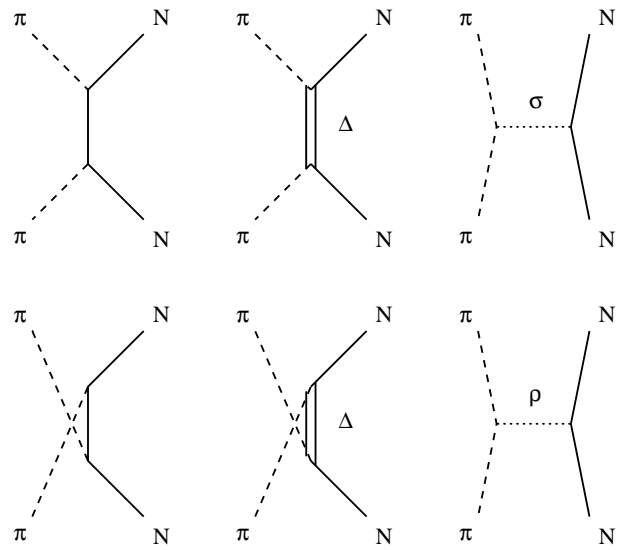,width=8cm}
\end{picture}
\caption[The Pion-Nucleon Interaction]
{Feynman-diagrams for the $\pi$N-interaction.}
\label{pinfd}
\end{figure}

Using the Lagrangians in Eqs.\ 
(\ref{smbb})-(\ref{rpp}), the
diagrams in Fig.\ \ref{pinfd} can be evaluated using standard
Feynman rules, attaching a factor $1/\sqrt{D_k^{(\pi)}}$ (Eq.\
(\ref{dk}) with $N_{pn}^{(\pi)}=3$) to each vertex with an external 
pion of momentum $k$, and replacing the standard Feynman propagator in
the $\sigma$- and $\rho$-exchange diagrams by the solitary meson
propagator, Eq.\ (\ref{pprop}) for the $\sigma$ and $\rho$ mesons,
respectively. To describe a self-interacting pion in the intermediate
state one has to modify the two-particle propagator of the 
Bethe-Salpeter (BS) scattering equation
\begin{equation}
\label{bse}
{\cal T}(p_{\mu}^{'},p_{\mu},s)={\cal V}(p^{'}_{\mu},p_{\mu},s)
+\int\frac{d^4q}{(2\pi)^4}
\;{\cal V}(p^{'}_{\mu},q_{\mu},s)
{\cal G}(q_{\mu},s){\cal T}(q_{\mu},p_{\mu},s),
\end{equation}
where $p_{\mu}$, $q_{\mu}$ and $p_{\mu}^{'}$ are the momenta of the
incoming, intermediate and outgoing nucleon, respectively. The
incoming particles are on their mass shell, i.\,e.\
$$
p_0=\sqrt{\vec{p}\,^2+M^2}\equiv\epsilon_N\qquad\mbox{and}\qquad
k_0=\sqrt{\vec{k}\,^2+m_{\pi}^2}\equiv\epsilon_{\pi}.
$$
In the center of mass (c.\,m.) system one gets
$$
s=(p_{\mu}+k_{\mu})^2=(p^{'}_{\mu}+k^{'}_{\mu})^2=
(\epsilon_N+\epsilon_{\pi})^2,
$$
and the pion momenta will be omitted since
$$
k_{\mu}=(\sqrt{s}-p_0,-\vec{p}\,)\qquad\mbox{and}\qquad
k^{'}_{\mu}=(\sqrt{s}-p_0^{'},-\vec{p}\,^{'}).
$$
The BS propagator then becomes 
\begin{equation}
\label{bspr}
{\cal G}(q_{\mu},s)=iP_{\pi}(p_{\mu}+k_{\mu}-q_{\mu})S_F(q_{\mu}).
\end{equation}
It is important to note that in Eq.\ (\ref{bspr}) the solitary meson
propagator is used for the intermediate pions instead of the Feynman
propagator. Due to the proper normalization, 
$iP_{\pi}(k_{\mu})$ now carries
{\it by construction} a sufficiently strong decay with 
increasing momentum
to regularize all diagrams so that phenomenological 
form factors are not needed. 
   
In the model of Pearce and Jennings \cite{PJ91}, there are two different
reduction schemes for the four-dimensional equation, Eq.\ (\ref{bse}). 
We use the `smooth-propagator' formalism since it has the
correct one-body limit \cite{CJ89}. The Blankenbecler-Sugar reduction
does not have this property. While this is not a major problem for
equal-mass systems such as NN scattering, it may cause problems in a
study of $\pi$N scattering. In conventional models, the reduction is 
performed using the substitution \cite{PJ91}, 
$$
i\Delta_F(p_{\mu}+k_{\mu}-q_{\mu})S_F(q_{\mu})
\quad\to\quad\delta(q_0-\epsilon_N)\;{\cal
G}_{sm}^{lin}(\vec{q},s),
$$
where
\begin{equation}
\label{smprlin}
{\cal G}_{sm}^{lin}(\vec{q},s)=\frac{2\pi}{\sqrt{s}}
\frac{\gamma_0\epsilon_N-\vec{\gamma}\vec{q}+M}
{\vec{p}\,^2-\vec{q}\,^2+i\epsilon}\ .
\end{equation}
This propagator is transformed to describe solitary mesons 
simply by setting
$$
iP_{\pi}(k_{\mu})\equiv i\Delta_F(k_{\mu})F_{\pi}(k_{\mu})=
\frac{i}{k^2-m_{\pi}^2}F_{\pi}(k_{\mu})\ ,
$$
and from Eq.\ (\ref{pprop}), one gets
\begin{equation}
\label{fpi}
F_{\pi}(k_0;|\vec{k}\,|)=\sum_{n=0}^{\infty} 
\frac{(m_{\pi}\alpha_{\pi})^{2n}
(2n+1)^{2n-2}}{{D_{k,n}^{(\pi)}}^{2pn+1}
\Big(\vec{k}\,^2+(2n+1)^2m_{\pi}^2\Big)^{n}}\;
\frac{k^2-m_{\pi}^2}{k^2-(2n+1)^2m_{\pi}^2}\ .
\end{equation}
Recall that the proper normalization constant was designed to yield in 
$iP_{\pi}(k_{\mu})$, a pole with residue $i$ at
$k^2=m_{\pi}^2$. Thus 
$F_{\pi}(k_0;|\vec{k}\,|)=1$ at the pion pole. 
The reduction of the Bethe-Salpeter
equation, Eq.\ (\ref{bse}), for solitary mesons can now be performed in
analogy to the development of Eq.\ (\ref{smprlin}) by the substitution 
$$
iP_{\pi}(p_{\mu}+k_{\mu}-q_{\mu})S_F(q_{\mu})\quad\to\quad 
\delta(q_0-\epsilon_N)\;{\cal G}_{sm}(\vec{q},s),
$$
where
\begin{equation}
\label{smpr}
{\cal G}_{sm}(\vec{q},s)=
\frac{2\pi}{\sqrt{s}}F_{\pi}(\epsilon_{\pi};|\vec{q}\,|)
\frac{\gamma_0\epsilon_N-\vec{\gamma}\vec{q}+M}
{\vec{p}\,^2-\vec{q}\,^2+i\epsilon}\ .
\end{equation}
Inserting Eq.\ (\ref{smpr}) into the Bethe-Salpeter
equation, Eq.\ (\ref{bse}) and performing a partial wave decomposition 
\cite{PJ91}, the one-dimensional scattering equation 
for the partial wave $T$-matrix ($p$ denotes $|\vec{p}\,|$ and $\ell$
stands for $\{L,T,J\}$)
\begin{equation}
\label{tmeq}
T_{\ell}(p^{'},p,s)=V_{\ell}(p^{'},p,s)+\int_0^{\infty}q^2dq\;
V_{\ell}(p^{'},q,s)\;G_{sm}(q,s)\;T_{\ell}(q,p,s),
\end{equation}
results, where
\begin{equation}
\label{gsm}
G_{sm}(q,s)=\frac{M}{(2\pi)^3\sqrt{s}}
\frac{F_{\pi}(\epsilon_{\pi};q)}{p^2-q^2+i\epsilon}\ .
\end{equation}
Explicit forms for the pseudopotentials, $V_{\ell}$, 
corresponding to the
Feynman amplitudes in Fig.\ \ref{pinfd}, evaluated with 
the model of Pearce and Jennings, are listed in Ref.\ \cite{PJ91}. 
The OSBEP pseudopotentials then are
obtained by replacing the form factors with 
$1/\sqrt{D_k^{(\pi)}}$ for each pion leg of momentum $k^{\mu}$ and
by substituting the Feynman propagators with the solitary meson
propagators in the $\sigma$- and $\rho$-exchange amplitudes.
Phase shifts are then calculated from the on-shell $T$-matrix on
defining the density of states by 
$$
\mbox{disc}\,G_{sm}(q,s)
=-\frac{2\pi i}{p^2}\rho(p)\delta(p-q),
$$
and with Eq.\ (\ref{gsm}) to have
$$
\rho(p)=\frac{pM}{(2\pi)^32\sqrt{s}}F_{\pi}(\epsilon_{\pi};q),
$$
so that defining 
$$
\tau_{\ell}(p)=-\pi\rho(p)T_{\ell}(p,p)
$$
the phase shifts can be specified by 
$$
\delta_{\ell}(p)=\arctan
\frac{\mbox{Im\,}\tau_{\ell}(p)}{\mbox{Re\,}\tau_{\ell}(p)}.
$$
\section{Results}
\label{res}
We calculated the NN and $\pi$N phase shifts separately and
compared the results with the latest single-energy phase shift
analyses; SM97 \cite{sm97} for NN and SM95 \cite{sm95} for $\pi$N 
scattering, respectively. Since there are no phenomenological 
form factors in our model and the scaling law relates all meson
nonlinearities to the pion self-interaction coupling constant
$\alpha_{\pi}$, that constant and the meson-baryon and
meson-meson coupling constants were the only parameters we adjusted to
achieve fits to data. Of these parameters, the 
tensor-vector ratio $\kappa$ and the pion self-interaction coupling
constant $\alpha_{\pi}$ are involved with both potentials. Hence those
two play a crucial role in the
determination of our optimal parameter set of values. 
We noticed that, when the $\pi$N data alone are considered, 
a rather low value of $\alpha_{\pi}$ (around 0.4) is favored. Alone, 
the NN system is much better described with a value of $\alpha_{\pi}$ of
about 0.7. However, this larger value can be reconciled with the $\pi$N
data. To do so one must set the value of $\kappa$ as low as possible
without losing much accuracy in fits to the NN data. 
   
We emphasize a good fit of the NN phase shifts as they are 
determined more accurate than are the $\pi$N phases and stay from a
larger database. Therefore, first we adjusted the parameters of the
model to find a fit to the NN data. It turned out to be 
even better than in our earlier work \cite{jae98}. Then,
we used the remaining parameters in a $\pi$N analysis 
to perform a fit with respect
to the SM95 phase shift analysis \cite{sm95}. We used those in
preference to the Karlsruhe-Helsinki phases \cite{kh80} as the SM95 data
have associated error bars which allow us to make a weighted fit. 
The ultimate parameter set values are listed in Table \ref{coco}. 
From those values note that the $\pi$NN 
coupling constant is smaller than the value of 
$g_{\pi}^2/{4\pi}=14.4$ previously used. The first indication that such
should be so came from
a Nijmegen analysis \cite{nigpi} which suggests $f_{\pi NN}^2=0.0745$ 
and thus $g_{\pi}^2/{4\pi}=13.79$ when our values for the 
pion and nucleon masses are used. Also, Arndt and 
co-workers with their analysis of $\pi$N scattering \cite{gpi} have
deduced a similar value. We confirmed that Arndt result in an
independent analysis \cite{SAN97} and so we fixed the   
$\pi$NN coupling constant to that value viz.\ 
\begin{equation}
\label{pinn}
\frac{g_{\pi}^2}{4\pi} = 13.75.
\end{equation}
The $\pi N\Delta$ coupling constant is then fixed by the
quark-model relation \cite{BW75}
$$
\frac{g_{\pi
N\Delta}^2}{4\pi}=\frac{72}{25}\left(\frac{m_{\pi}}{2M}\right)^2
\frac{g_{\pi}^2}{4\pi}\ .
$$
It should be noted that the large value of the pion self-interaction
coupling constant ($\sim 0.7$) can only be used in the $\pi$N potential
if the $\pi N\Delta$ coupling constant is set to that quark model value.
If one uses the value $g_{\pi N\Delta}^2/{4\pi}=0.36$, 
as chosen for most
other $\pi$N potentials, the fit is much worse. 
Another important feature in Table \ref{coco} is the sign of the 
$\sigma\pi\pi$ coupling constant. In the work of Pearce and Jennings
\cite{PJ91}, this coupling is positive and very large
($g_{\sigma\pi\pi}g_{\sigma}/{4\pi}=143.6$), which may be 
caused by the rather low cut-off mass ($\sim 500$\,MeV) 
they use in the form factor of the
$\sigma$NN and $\sigma\pi\pi$ vertices. Such a cut-off is very abrupt. 
Furthermore, using a model based on correlated
two-pion exchange derived from dispersion relations, Sch\"utz {\it et
al.} \cite{SDHS94} found the sign of the product 
$g_{\sigma\pi\pi}g_{\sigma}$ should be negative.
  
Since the $\Delta$ and nucleon pole diagrams are iterated in the $\pi$N
scattering equation, one has to use the bare values for masses and
coupling constants in the kernel of the integral equation for the
$P_{33}$ and $P_{11}$ channels, respectively. In principle, these values
are related to the physical ones by the renormalization procedure 
\cite{PA86}. We simplified the model by finding the bare values that
optimize the fit to the 
phase shifts in the relevant channels. First, we adjusted the
other parameters to fit the phase shifts in the non-resonant channels. 
After that, there was but 
one bare mass and coupling constant for the nucleon and $\Delta$ 
which reproduced the phase shifts in the $P_{11}$ and
$P_{33}$ channels, respectively. 
By this procedure, in principle the bare parameters were functions
of the other parameters, too. The results are given in Table \ref{ren}. 
The values of the parameters in Table \ref{coco} involved with the NN
potential are very similar to those found using our original (pure NN)
potential  \cite{jae98}. The proper
normalizations of the $\pi$ and $\eta$ 
are the only features that vary, it is not 
surprising that the only significant change in the parameter values is
that for the $\eta$NN coupling constant, the present result being
considerably less than the former value of 0.702 \cite{jae98} and that
for the self-interaction coupling
constant, $\alpha_{\pi}$, the present result being much larger than
found with the fit using $N_{pn}^{(ps)}=1$ (there $\alpha_{\pi}=0.44065$
\cite{jae98}). However, the generalized scaling law, Eq.\
(\ref{scaling}), keeps the vector and scalar self-interaction coupling 
constants close to the values determined by the older fit.
   
We have used OSBEP to fit to fit NN phase shifts (to 300\,MeV) for
numerous angular momentum channels and to fit $\pi$N phase shift data in
all $S$ and $P$ channels to a momentum of 500\,MeV/$c$. Excellent fits
have been obtained as is evident from Table \ref{chisq} in which the
$\chi^2$/datum with respect to the world's NN database are listed in
comparison to those found with standard models. A byproduct is that
OSBEP yields excellent results for the properties of the deuteron. They
are listed in Table \ref{deut} wherein comparison is made with the
experimental values and with those associated with the Bonn-B force. 
  
\begin{figure}[t]\centering
\begin{picture}(14.0,20.0)(0.0,-0.5)
\epsfig{figure=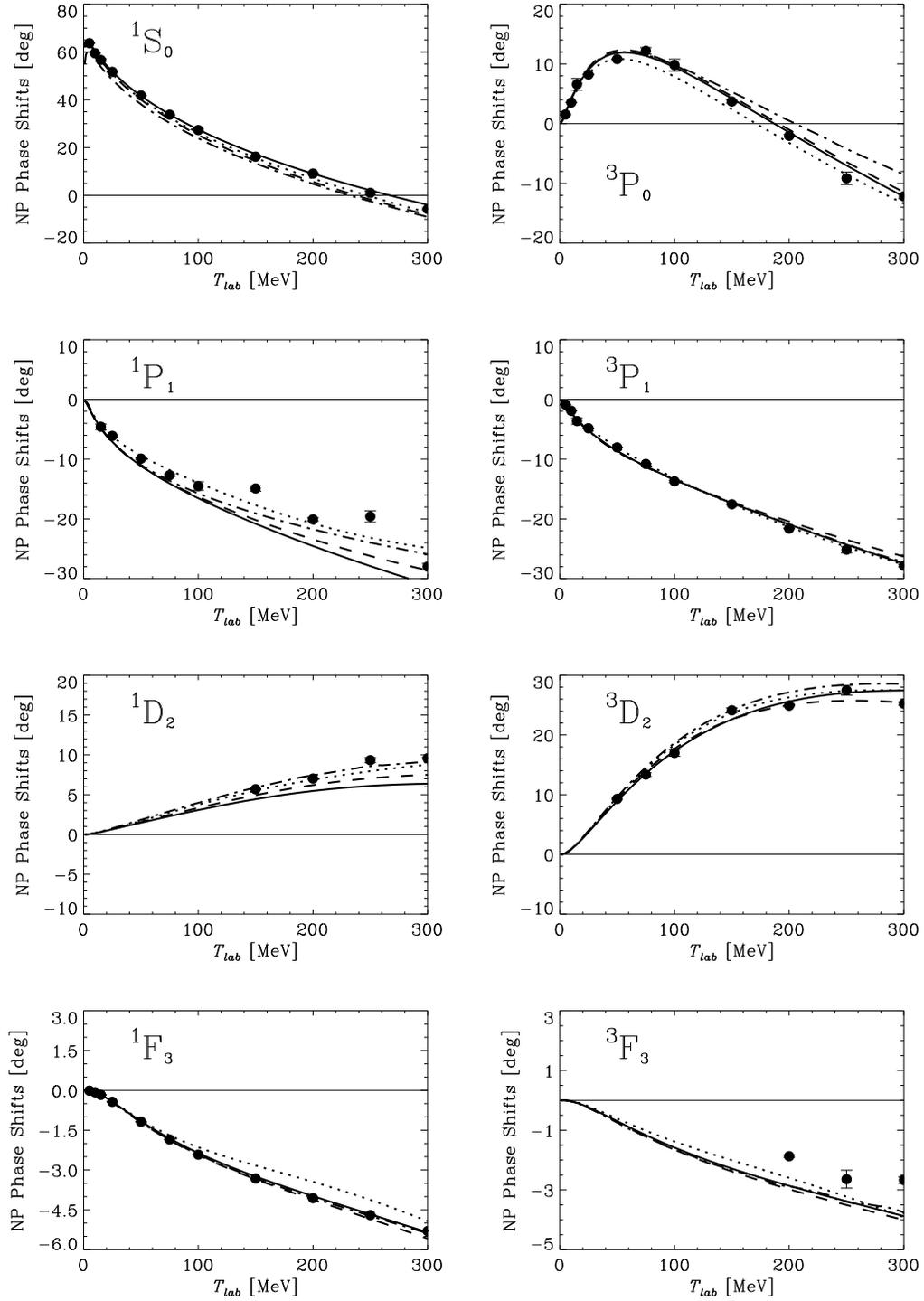,width=14.0cm}
\end{picture}
\caption[$np$ Phase Shifts]{$np$ phase 
shifts. The Arndt SM97 \cite{said} phase shifts
(circles) are compared with the phase shifts calculated using the Nijm93
\cite{nipot} (dotted), Bonn-B \cite{Mach89} (dashed), Paris
\cite{papot} (dash-dotted) potentials and with our OSBEP (solid).} 
\label{npph}
\end{figure}

\begin{figure} \centering
\begin{picture}(16.0,20.0)(0.0,0.0)
\epsfig{figure=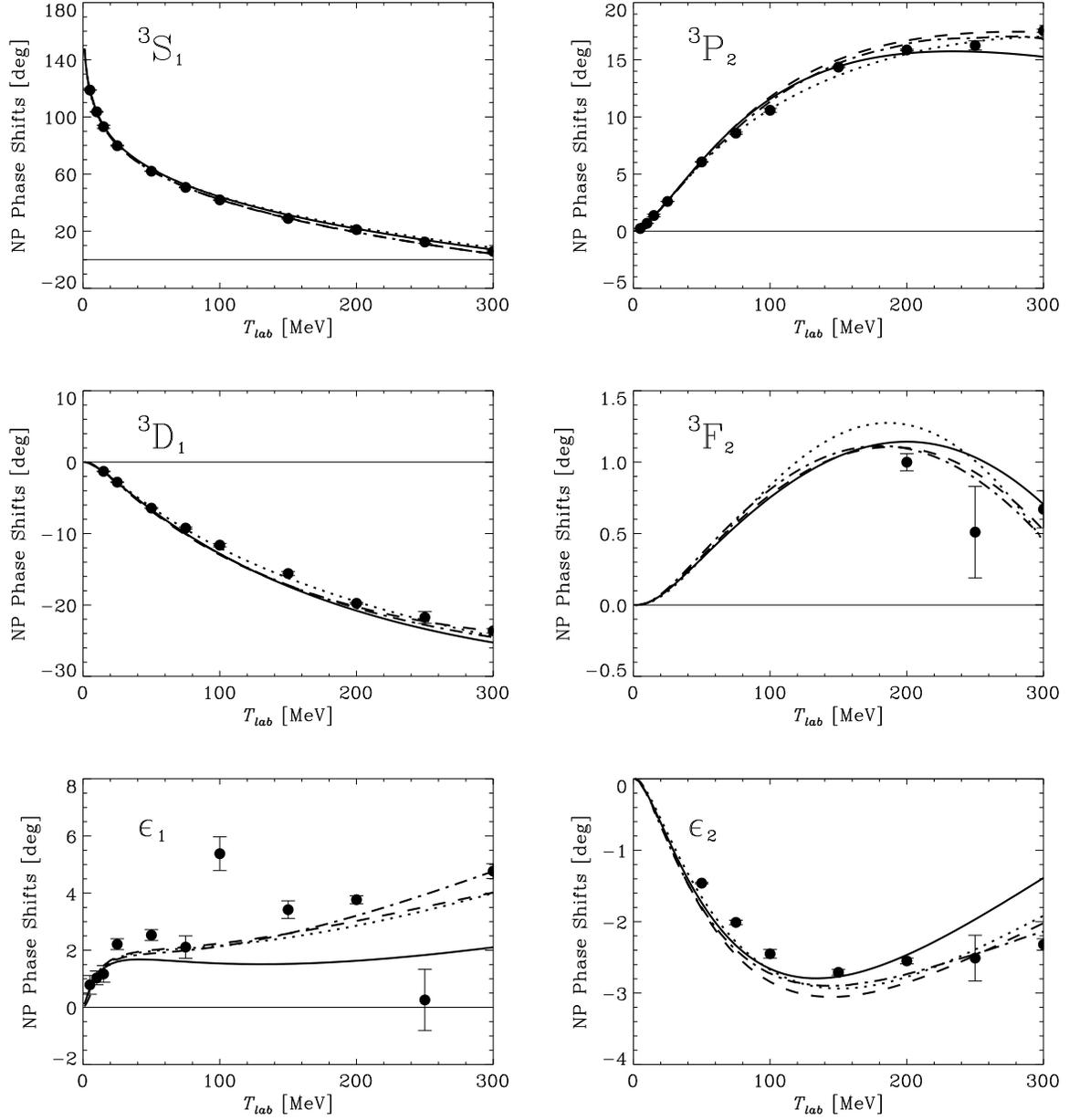,width=16.0cm}
\end{picture}
\caption{SYM $np$ phase shifts for the coupled $^3SD_1$ and 
$^3PF_2$ channels with notation as in Figure\,\protect\ref{npph}.
\label{coupl}}
\end{figure}

\begin{figure}[t]\centering
\begin{picture}(14.0,20.0)(0.0,-0.5)
\epsfig{figure=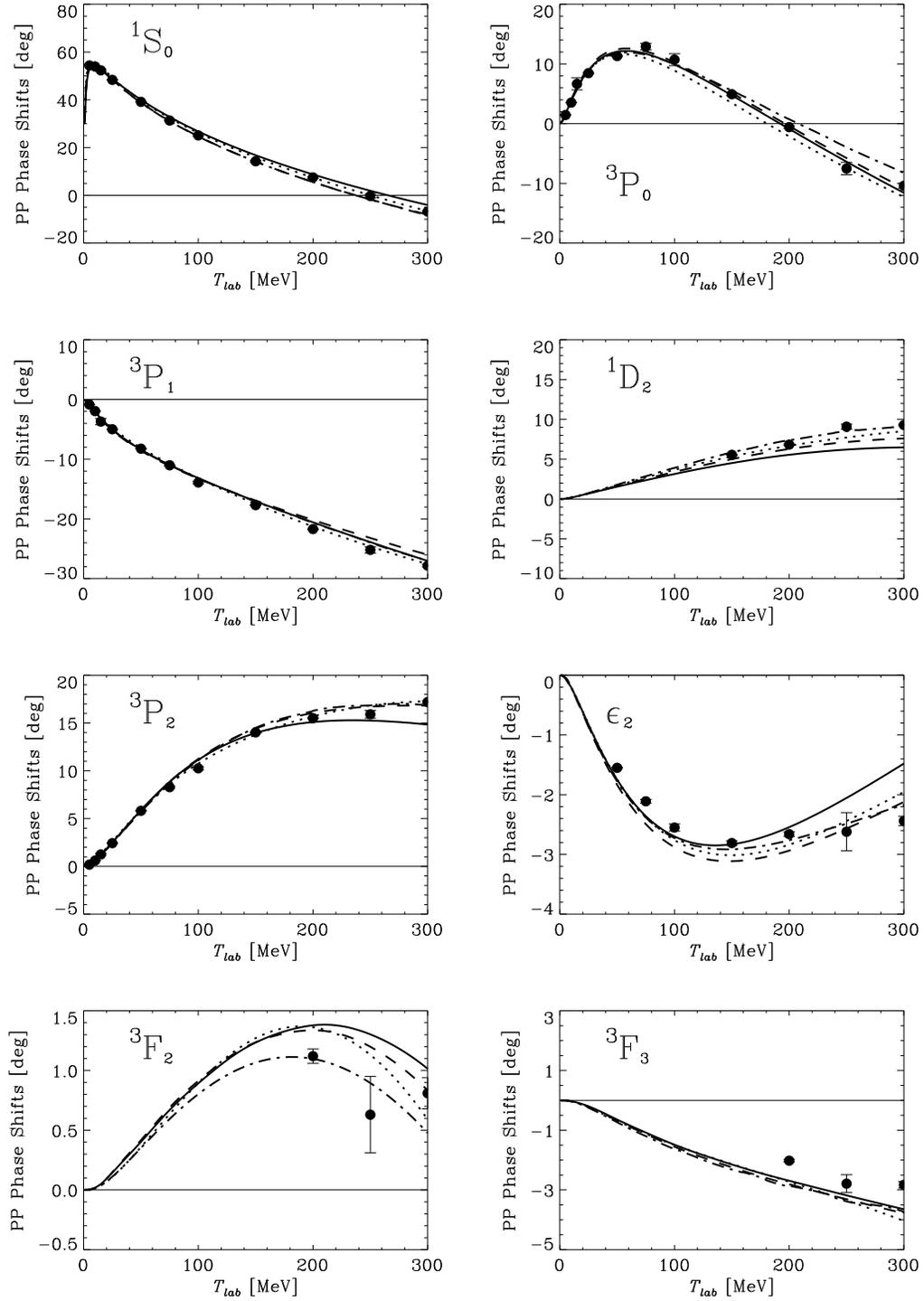,width=14.0cm}
\end{picture}
\caption[$pp$ Phase Shifts]{SYM $pp$ phase 
shifts. The Arndt SM97 \cite{said} phase shifts
(circles) are compared with the results of calculations made using the
Nijm93 \protect\cite{nipot} (dotted), Bonn-B (dashed, see 
\protect\cite{jae98}), Paris \protect\cite{papot} 
(dash-dotted) potentials and with our OSBEP (solid).}
\label{ppph}
\end{figure}

\begin{figure}\centering
\begin{picture}(16,20)(0.0,0.0)
\epsfig{figure=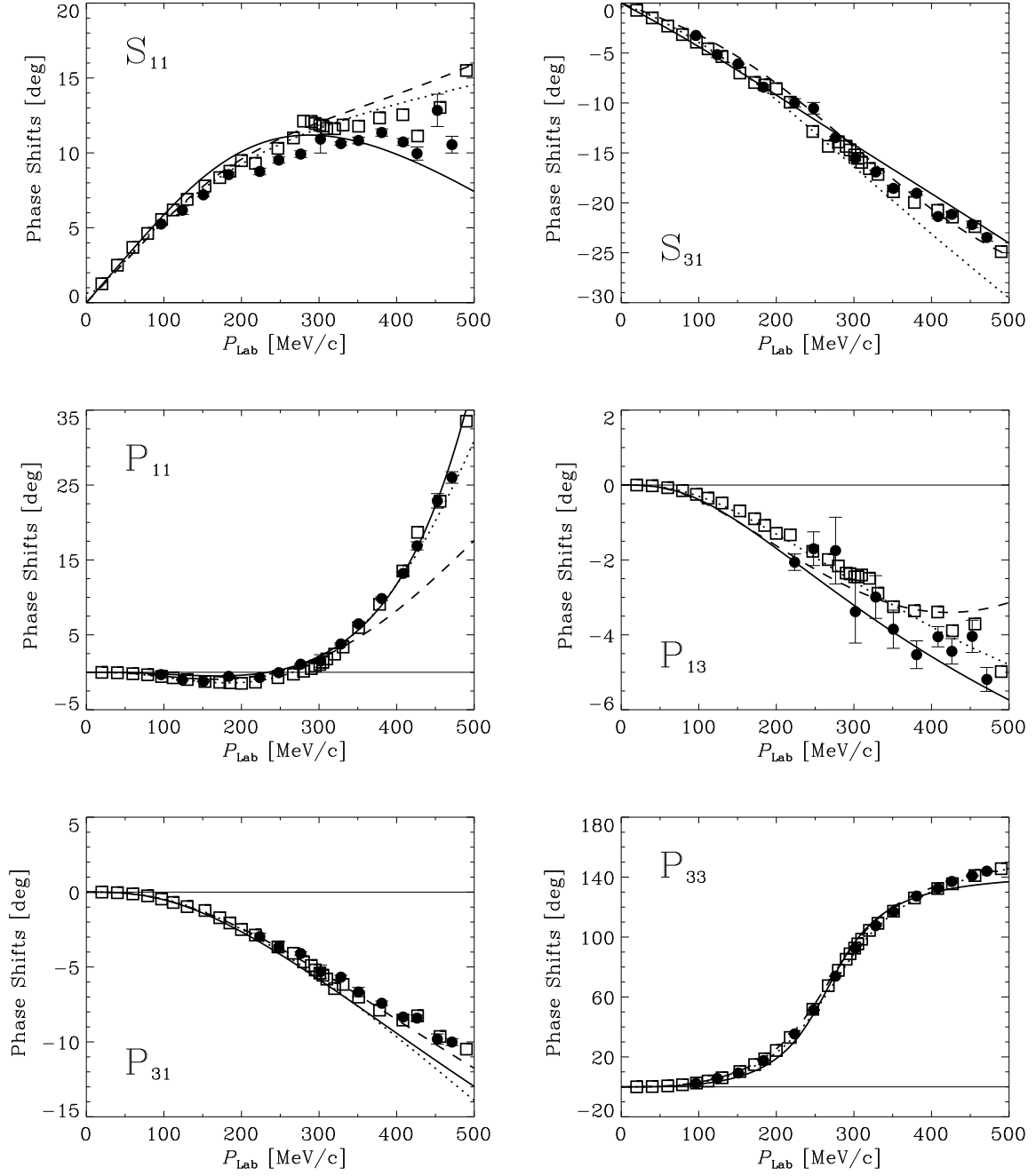,width=16cm}
\end{picture}
\caption{$\pi N$ phase shifts. The SM95 \protect\cite{sm95}
(dots) and KH80 \protect\cite{kh80} (squares)
phase shift analyses compared with results of calculations made by
Pearce and Jennings \protect\cite{PJ91} (dashed), Sch\"utz {\it et al.}
\protect\cite{SDHS94} (dotted) and with our OSBEP model (solid).}
\label{pinph}
\end{figure}

The phase shifts for diverse channels are compared with data and the
predictions of standard models in Figs.\ \ref{npph}-\ref{ppph} for NN
scattering and in Fig.\ \ref{pinph} for $\pi$N scattering. In Fig.\
\ref{npph} the $np$ phase shifts for uncoupled channels (to $^3F_3$) are
shown. The OSBEP results are as good if not better than those of the
standard models with rare exception. That is also the case with the
coupled channels in Fig.\ \ref{coupl}. Finally, in Fig.\ \ref{ppph}, we
show the $pp$ phase shifts to which OSBEP does as well as the
conventional potential calculations.
  
The $S$ and $P$ wave channel phase shifts for $\pi$N scattering as given
by OSBEP and two other model calculations are compared with data in
Fig.\ \ref{pinph}. The OSBEP results are again good and of a quality
comparable to that found with the other model results.  
   
However, while the OSBEP fit to the NN data is very satisfactory, 
providing at least the same quality as conventional models with a
minimum number of the parameters, the $\pi$N fits 
could be further improved. Especially, in the $S_{11}$ channel, 
inclusion of the $N^{\ast}(1535)$ resonance would contribute by
increasing the value of the phase shifts at energies above 400\,MeV
\cite{S95}. We note also that the width of the $\Delta$ resonance in the
$P_{33}$ channel predicted by OSBEP is not as accurate as those found
with the conventional models. The resonance is produced mainly from the
background potential and not just from the $\Delta$ pole diagram
alone and which reflects in the rather low value of the bare $\Delta$
mass listed in Table \ref{ren}. At the same time, the background
potential has to compensate for the negligible effect of the
$N^{\ast}(1535)$ in the $S_{11}$ channel phase shifts. 
Inclusion of this resonance in the model would
simultaneously improve the fit to the $P_{33}$ channel data. 
Note that the $\pi$N data fit was performed using the
SM95 phase shifts (dots) with their error bars as experimental input to
a search. Thus, the OSBEP phases must deviate from the KH80 phase shifts
values (squares) in the $P_{13}$ channel.
\subsection{Soft-pion theorems}
\begin{figure}[t] \centering
\begin{picture}(8,8)(0.0,0.0)
\epsfig{figure=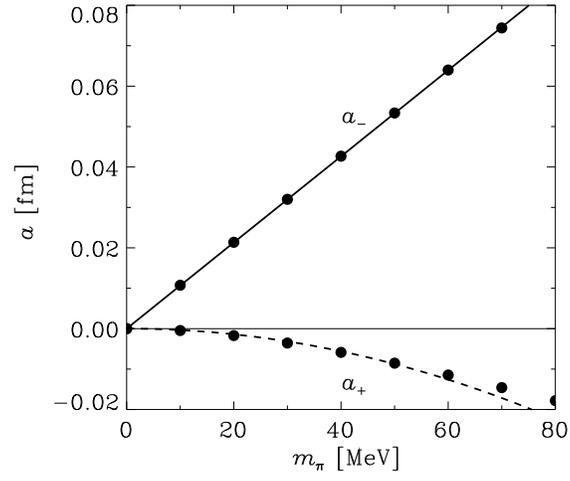,width=8cm}
\end{picture}
\caption[The Pion-Nucleon Interaction]
{$S$-wave scattering lengths from $\pi$N scattering calculations made
using OSBEP as a function of the pion mass. 
\label{sl}}
\end{figure}

To test whether the model restores chiral symmetry in the limit
$m_{\pi}\to 0$, we calculate the $S$-wave scattering lengths and
compare them with the Weinberg-Tomozawa relations,
$$
a_+ = \frac{1}{3}(a_{S_{11}}+2a_{S_{31}})={\cal O}(m_{\pi}^2),
$$
and
$$
a_- = \frac{1}{3}(a_{S_{11}}-a_{S_{31}})={\cal O}(m_{\pi}).
$$
derived from the soft-pion theorems. 
The scattering lengths are plotted as a function of the pion mass in
Fig.\ \ref{sl}. It is obvious that both slopes follow the
Weinberg-Tomozawa relations nicely and thus the model
of solitary mesons does not contradict the soft-pion theorems.
\section{Summary and outlook}
\label{sum}
In this work we have shown that the one solitary boson exchange
potential OSBEP can be extended to describe simultaneously NN and $\pi$N
scattering data. With this approach, we have no problem in having a
consistent description of both systems. There is no 
incompatibility of the $\pi$NN form factor in particular. Since our 
solitary boson exchange method regularizes the self-energy diagrams 
{\it a priori}, the model enabled us to obtain 
consistently finite scattering amplitudes for NN
as well as $\pi$N scattering. 
Additionally, we were able to retain the empirical scaling
relation which already was successfully applied in a precise analysis of
NN scattering alone. This relation serves to significantly reduce the
number of parameters existent in our model below that required with all
other methods. The model phase shifts agree very well with those found
using the latest NN and $\pi$N phase
shift analyses and, with the properties of the deuteron. The accuracy of
the fits are comparable to those given by conventional potential models
for NN and $\pi$N respectively. 
  
In future we hope to apply this model in analyses of pion production
processes. It is well known that a proper description of the very
accurate data near threshold demands a NN final state interaction as
well as a $\pi$N $T$-matrix that are consistent with each
other. The solitary boson exchange potential fulfills
this need. Use of OSBEP to analyze 
$\pi\pi$ scattering is another interesting aim.
It would be a serious test for this model to see if the dynamics of
solitary mesons are compatible with such data and if the model can
maintain the consistency we have found by studying 
the NN and $\pi$N systems.   

Finally we note a need to perform a refined simultaneous fit to NN and
$\pi$N and the calculation of $\pi$N scattering observables.
Since the simultaneous fit to NN and $\pi$N data is very 
time-consuming, the phase shifts 
shown here were obtained first by fitting the NN data and then by
adjusting the remaining three parameters to fit the $\pi$N data. 
Therefore, the quality of fit to the NN phases is better than that to
the $\pi$N ones. However, the accuracy of our results convince us that
the solitary boson exchange potential works consistently for NN and
$\pi$N interactions.
\acknowledgements
The author would like to thank K. Amos from the University of Melbourne
for intensive review of the manuscript. This work was
supported in part by Forschungszentrum J\"ulich GmbH under Grant No.\
41126865. 
\tighten
%
\begin{table} 
\caption{Meson masses and proper normalizations associated with OSBEP.}
\label{mm}
\begin{tabular}{lcccccc}
 $\beta$ & $\pi$ & $\eta$ & $\rho$ & $\omega$ & $\sigma$ & $\delta$\\
\hline
$m_{\beta}\mbox{ [MeV]}$ & 138.03\tablenotemark[1] & 548.8 & 769 &
782.6 & 550\tablenotemark[2] & 983 \\
$N^{(\beta)}_{pn}$ & 3 & 3 & 2 & 2 & 1 & 1 \\ 
\end{tabular}
\tablenotetext[1]{For the $pp$ potential, we used the neutral pion mass
$m_{\pi_0}=134.9764$\,MeV.} 
\tablenotetext[2]{For the $T=0$ $np$ potential, we adopted 
$m_{\sigma}=720$\,MeV from the Bonn-B potential.} 
\end{table}
%
%
\begin{table} 
\caption{The optimal parameter values of our OSBEP model. The parameters
influencing phase shift calculations for the NN and $\pi$N scattering
systems are indicated.}
\label{coco}
\begin{tabular}{clcc}
 Name & Value & NN & $\pi$N \\
\hline
$g_{\pi      }^2/{4\pi}$ & 13.75 (fixed) & x & x \\
$\alpha_{\pi}$           & 0.7471        & x & x \\
$\kappa_{\rho}$          & 3.3982        & x & x \\
$g_{\eta     }^2/{4\pi}$ & 0.0745        & x &   \\
$g_{\rho     }^2/{4\pi}$ & 1.6725        & x &   \\
$g_{\omega   }^2/{4\pi}$ & 22.499        & x &   \\
$g_{\sigma_0 }^2/{4\pi}$ & 12.2415       & x &   \\
$g_{\sigma_1 }^2/{4\pi}$ & 8.9523 (np)   & x &   \\
$g_{\sigma_1 }^2/{4\pi}$ & 8.8461 (pp)   & x &   \\
$g_{\delta   }^2/{4\pi}$ & 1.4172        & x &   \\
$g_{\rho\pi\pi}g_{\rho}/{4\pi}$           &   5.7047 &   & x \\
$g_{\sigma\pi\pi}g_{\sigma}/{4\pi}$ & -0.7434 &   & x \\
$g_{\pi N\Delta}^2/{4\pi}$                &  0.213954 (fixed)&   & x \\
$x_{\Delta}$                              & -0.1829 &   & x \\
\end{tabular} 
\end{table}
%
%
\begin{table} 
\caption{Bare and renormalized values for nucleon and $\Delta$ masses
and coupling constants. The bare values are used in the pseudopotentials
for the nucleon and $\Delta$ pole diagrams in the $P_{11}$ and $P_{33}$
channel, respectively.}
\label{ren}
\begin{tabular}{lllll}
 & $M$ [MeV] & $g_{\pi}^2/{4\pi}$ & $M_{\Delta}$ [MeV] 
& $g_{\pi N\Delta}^2/{4\pi}$ \\
\hline
bare & 1346.51 & 1.8687 & 1027.80 & 0.0437 \\
dressed & 938.926 & 13.75 & 1232 & 0.2139 \\
\end{tabular}
\end{table}
%
%
\begin{table} 
\caption{$\chi^2$/datum for the OSBEP and several potential models. Data
and $\chi^2$ values for the Nijm93 and Paris 
potential were taken from {\sc SAID} \protect\cite{said}.}
\label{chisq}
\begin{tabular}{lllll}
\multicolumn{1}{c}{Model}  & \multicolumn{1}{c}{No.\ of param.} & 
\multicolumn{1}{c}{$np$\tablenotemark[1]}
& \multicolumn{1}{c}{$pp$\tablenotemark[2]} & 
\multicolumn{1}{c}{Total} \\\hline
 OSBEP        & 8  & 2.9 & 6.7 & 4.1 \\
 Nijm93 & 15 & 5.6 & 2.2 & 4.5 \\
 Bonn-B      & 15 &12.1 & 5.8\tablenotemark[3] & 10.1 \\
 Paris        &$\approx$60 & 12.5 & 2.3 &  9.2 \\
\end{tabular}
\tablenotetext[1]{Energy bin 1-300\,MeV (2713 data points).}
\tablenotetext[2]{Energy bin 1-300\,MeV (1292 data points).}
\tablenotetext[3]{$pp$ version
$g_{\sigma_1}^2/{4\pi}=8.8235$, see \protect\cite{jae98}.}
\end{table}
%
%
\begin{table} 
\caption{The properties of the deuteron.}
\label{deut}
\begin{tabular}{lllll}
 & \multicolumn{1}{c}{Bonn-B \protect\cite{Mach89}} &
\multicolumn{1}{c}{OSBEP} & 
\multicolumn{1}{c}{Exp.} &
\multicolumn{1}{c}{Ref.} \\
\hline
$E_B\mbox{ [MeV]}$ & $2.2246$ &    $2.22459$   
&  $2.22458900(22)$ &
\protect\cite{red}  \\   
$ \mu_d$           & $0.8514$\tablenotemark[1]  &  
$0.8456$\tablenotemark[1] &  $0.857406(1)$    & 
\protect\cite{rmu}  \\
$Q_d\mbox{ [fm$^2$]}$ &  $0.2783$\tablenotemark[1] &
$0.2728$\tablenotemark[1] &  $0.2859(3)$      &
\protect\cite{rqa}   \\
$A_S\mbox{ [fm$^{-1/2}$]}$  &  $0.8860$    &
   $0.8788$    &   $0.8802(20)$  &
\protect\cite{rqa}   \\
$D/S$           &   $0.0264$    &           
   $0.0256$    &   $0.0256(4)$   &
\protect\cite{rds}   \\
$r_{RMS}\mbox{ [fm]}$ &  $1.9688$     &      
  $1.9554$     &  $1.9627(38)$      &
\protect\cite{rqa}   \\
$P_D\quad[\%]$     &   $4.99$      &          
   $6.00$      &
\multicolumn{1}{c}{-}    
&   \multicolumn{1}{c}{-}  \\
\end{tabular}
\tablenotetext[1]{Meson exchange current contributions not 
included}    
\end{table}
%
\clearpage
%
%
%
%
%
%
%
%
%
%
%
%
%
%
%
%
%
%
\end{document}